# Effects of Pore-scale on the Macroscopic Properties of Natural Convection in Porous Media[†]

Stefan Gasow[1], Zhe Lin[2], Hao Chun Zhang[3], Andrey V. Kuznetsov[4], Marc Avila[1], and Yan Jin[1*]

[1] *Center of Applied Space Technology and Microgravity (ZARM), University of Bremen, 28359, Bremen*

[2] *Key Laboratory of Fluid Transmission Technology of Zhejiang Province, Zhejiang Sci-Tech University, 5 Second Avenue, Xiasha Higher Education Zone, Hangzhou 310018, China*

[3] *School Energy Science and Engineering, Harbin Institute of Technology, Harbin 150001, China*

[4] *Department of Mechanical and Aerospace Engineering, North Carolina State University, Raleigh, North Carolina 27695-7910, USA*

[*] *Corresponding author: Yan Jin (yan.jin@zarm.uni-bremen.de)*

**Abstract**

Natural convection in porous media is a fundamental process for the long-term storage of $CO_2$ in deep saline aquifers. Typically, details of mass transfer in porous media are inferred from the numerical solution of the volume-averaged Darcy-Oberbeck-Boussinesq (DOB) equations, even though these equations do not account for the microscopic properties of a porous medium. According to the DOB equations, natural convection in a porous medium is uniquely determined by the Rayleigh number. However, in contrast with experiments, DOB simulations yield a linear scaling of the Sherwood number with the Rayleigh number (Ra) for high values of Ra (Ra>>1,300). Here, we perform Direct Numerical Simulations (DNS), fully resolving the flow field within the pores. We show that the boundary layer thickness is determined by the pore size instead of the Rayleigh number, as previously assumed. The mega- and proto- plume sizes increase with the pore size. Our DNS results exhibit a nonlinear scaling of the Sherwood number at high porosity, and for the same Rayleigh number, higher Sherwood numbers are predicted by DNS at lower porosities. It can be concluded that the scaling of the Sherwood number depends on the porosity and the pore-scale parameters, which is consistent with experimental studies.

---

[†] Cite as: S. Gasow, Z. Lin, Z., H.C. Zhang, A.V. Kuznetsov, M. Avila, M., Y. Jin (2020) "Effects of pore-scale on the macroscopic properties of natural convection in porous media," *J. Fluid Mech.*, 891, A25.

**I. Introduction**

Since the seminal work of Horton & Rogers (1945) and Lapwood (1948), buoyancy-driven convection in a cell occupied by a porous medium has been adopted as a model for many applications. Some examples of these applications include oil recovery, ground water flow, and geothermal energy extraction. Convection in porous media may be caused by heat or mass transfer. The key governing parameter is the Rayleigh-Darcy number (Ra-Da), which is more often referred to as the Rayleigh number (Ra). It is a combination of a traditional Rayleigh number $Ra_f$ and a Darcy number Da. A traditional Rayleigh number describes the ratio of buoyancy to viscosity within a fluid, multiplied by the Prandtl number (for heat transfer) or Schmidt number (for mass transfer). A Darcy number describes the ratio between the permeability and the characteristic area. A porous medium produces two effects on convection. First, the largest length scale of fluid motions becomes limited by pore size, and thus heat/mass transfer is reduced (Jin et al. 2015, 2017; Uth et al. 2016). Second, the interaction between flow and porous elements may lead to dispersion processes within the pores, which can enhance heat/mass transfer (Nield & Bejan 2017).

In recent years, natural convection in porous media has received increasing attention due to the interest in the long-term storage of $CO_2$ in deep saline aquifers (Huppert & Neufeld 2014), where $CO_2$ sequestration is driven by gradients of $CO_2$ concentration. The efficiency of $CO_2$ sequestration is determined by the Sherwood number (Sh), which is the ratio of the total mass transfer rate (by convection and mass diffusion) to the diffusive mass transfer rate across a wall surface. Most reservoirs have small Rayleigh numbers (Ra < 1,500) (Hassanzadeh 2007), but some $CO_2$ reservoirs have a thickness of several hundred meters, and are thus characterized by very high Rayleigh numbers (Ra ≈ 12,000) (Neufeld et al., 2010).

Many experimental, theoretical, and numerical studies have been performed to determine the scaling $Sh = f(Ra)$. Theoretical and numerical studies rely on the Darcy-Oberbeck-Boussinesq (DOB) equations (Nield & Bejan, 2017), where the microscopic details of the porous media are solely accounted for by the permeability and effective diffusivity through Ra. This is explained by the loss of information during volume averaging.

Convection in porous media can be classified into five regimes based on the Rayleigh number (Nield & Bejan, 2017):

    I.    the conducting regime ($0 \leq Ra \leq 4\pi^2$);

II. the steady state regime ($4\pi^2 \leq \text{Ra} \leq 350$);
III. the quasi-periodic regime, which is dominated by two coherent convecting cells ($350 \leq \text{Ra} \leq 1{,}300$);
IV. the high Rayleigh regime ($1{,}300 \leq \text{Ra} \leq 10{,}000$), which is considered chaotic and "turbulent" without large coherent structures;
V. the ultimate Rayleigh regime ($\text{Ra} \geq 10{,}000$), which differs from the high Rayleigh regime only by the increasing self-organization of the inner flow field.

Howard (1964) suggested a linear scaling of Sh versus Ra at asymptotically large Rayleigh numbers. This was later proven in the analytical study of Doering & Constantin (1998). With the advent of high-performance computing, several numerical studies of the DOB equations (herein DOB simulations) for large Rayleigh numbers have been performed. The maximum Rayleigh numbers in these studies reach up to $\text{Ra} = O(10^4)$. For example, the DOB simulations of Otero *et al.* (2004) suggest scaling of the form $\text{Sh} \sim \text{Ra}^\alpha$ with $\alpha \approx 0.9$ for $\text{Ra} \leq 10{,}000$, which is slightly different from the theoretical linear scaling. However, subsequent DOB simulations, with Ra up to 40,000, confirmed the linear scaling of Sh with Ra for both isotropic (Hewitt *et al.* 2012, 2013, 2014; Wen *et al.* 2015) and anisotropic porous media (Paoli *et al.* 2016).

Despite this encouraging agreement between theory and numerical simulations, experimental measurements of mass transfer yielded lower exponents. Backhaus *et al.* (2011) and Neufeld *et al.* (2018) reported $\text{Sh} \sim \text{Ra}^{0.8}$. The experiments were also performed at very large Rayleigh numbers ($\text{Ra} \gg 1{,}300$). Hence, discrepancies occur between the theoretical value of the exponent ($\alpha = 1$) and the values of the exponent obtained from laboratory experiments.

It should be noted that many early numerical studies demonstrated nonlinear scaling of the Sherwood number, see Trevisan & Bejan (1987), Robinson & O'Sullivan (1976), and Caltagirone (1975). However, the maximum Rayleigh number simulated by Trevisan & Bejan (1987) and Caltagirone (1975) was around 2,000, which is only marginally larger than the transition Rayleigh number between regimes III and IV. Robinson & O'Sullivan (1976) assumed that the flow is steady. Therefore, the nonlinear relationship between Sh and Ra found in the earlier numerical studies cannot explain the discrepancies between numerical and experimental results observed at very large Rayleigh numbers.

These discrepancies cast doubt on the key hypothesis underlying the DOB equations, namely that the sole control parameter of convection in porous media is the Rayleigh number. This may result in serious model errors because of the many physical simplifications it carries. For example, Mijic *et al.* (2014) argued that the Forchheimer term (accounting for the quadratic drag typical of turbulence flows) may have an important effect on convection, whereas Wen *et al.* (2018) argued that convection in a porous medium is also influenced by the dispersion term. Thus, the fundamental question arises as to whether natural convection in porous media is governed by parameters other than Ra through additional physical mechanisms.

In this paper, we probe this question by performing pore-resolved Direct Numerical Simulations (DNS) of convection in porous media, thereby accounting for all scales of motion. We compare our DNS to traditional DOB simulations to determine whether parameters other than Ra influence convection in porous media.

## 2. Governing equations and numerical methods

We considered natural convection in a chamber filled with porous elements, see figure 1(a). The upper and lower walls were kept at species concentrations $c_1$ and $c_0$, respectively. The density difference caused by the different species concentrations at the lower and upper walls leads to natural convection in the chamber.

2.1 Governing equations for DNS

The governing equations for the DNS are the Navier-Stokes equations, with the buoyancy force accounted for by the Boussinesq approximation, and the species concentration equation. Using the Einstein's summation convention, these equations are:

$$\frac{\partial u_i}{\partial x_i} = 0, \tag{1}$$

$$\frac{\partial u_i}{\partial t} + \frac{\partial (u_i u_j)}{\partial x_j} = -\frac{\partial p}{\partial x_i} + \nu \frac{\partial^2 u_i}{\partial x_j^2} + \beta g_i (c - c_0), \tag{2}$$

$$\frac{\partial c}{\partial t} + \frac{\partial (u_i c)}{\partial x_i} = D_f \frac{\partial^2 c}{\partial x_j^2}, \tag{3}$$

where $\beta$ is the concentration expansion coefficient defined as $\beta = \beta(c_0) = -\frac{1}{\rho_0}\left(\frac{\partial \rho}{\partial c}\right)_{c_0}$, $g_i$ is the $i$th component of the gravitational vector, $\nu$ is the kinematic viscosity, and $D_f$ is the mass diffusivity of the species. The no-slip boundary condition is used at all solid surfaces, whereas the mass transfer rates at the surfaces of the solid obstacles are set to zero because they cannot be penetrated by $CO_2$.

Using the characteristic concentration difference $\Delta c = c_1 - c_0$, velocity $u_m = \beta \Delta c g K / \nu$, length $H$, and time $t_m = H/u_m$, we obtained the following dimensionless governing equations:

$$\frac{\partial \tilde{u}_i}{\partial \tilde{x}_i} = 0, \tag{4}$$

$$\frac{\partial \tilde{u}_i}{\partial \tilde{t}} + \frac{\partial (\tilde{u}_i \tilde{u}_j)}{\partial \tilde{x}_j} = -\frac{\partial \tilde{p}}{\partial \tilde{x}_i} + \frac{\text{Sc}}{\text{Ra}_f \text{Da}} \frac{\partial^2 \tilde{u}_i}{\partial \tilde{x}_j^2} - \frac{\text{Sc}}{\text{Ra}_f \text{Da}^2} z_i \tilde{c}, \tag{5}$$

$$\frac{\partial \tilde{c}}{\partial \tilde{t}} + \frac{\partial (\tilde{u}_i \tilde{c})}{\partial \tilde{x}_i} = \frac{1}{\text{Ra}_f \text{Da}} \frac{\partial^2 \tilde{c}}{\partial \tilde{x}_j^2}, \quad (6)$$

where $\tilde{c} = \frac{c - c_0}{c_1 - c_0}$ is the non-dimensional species concentration, $\text{Ra}_f = \frac{H^3 \beta \Delta c g}{\nu D_f}$ is the Rayleigh number for the free fluid flow between the bounded walls, $\text{Sc} = \frac{\nu}{D_f}$ is the Schmidt number, $\text{Da} = \frac{K}{H^2}$ is the Darcy number, $H$ is the height of the chamber, $g$ is gravity, $z_i$ is the $i$th component of the unit vector pointing in the direction of gravity, and $K$ is the permeability.

We calculated the Sherwood number from our DNS as the ratio of the total mass transfer rate $\dot{m}$ (by convection and diffusion) to diffusive mass transfer rate $\dot{m}_\text{diff}$ across the wall:

$$\text{Sh} = \dot{m}/\dot{m}_\text{diff} = \frac{\int_w \overline{\frac{\partial \tilde{c}}{\partial \tilde{x}_2}} dA}{\int_w \overline{\frac{\partial \tilde{c}}{\partial \tilde{x}_2}}\Big|_{\text{Ra}=0} dA}. \quad (7)$$

The subscript $w$ denotes either the lower or upper wall surface, and the overbar ¯ denotes the time average.

2.2 Governing equations for macroscopic simulations

We derived the governing equations for the macroscopic simulations using an approach similar to that used in de Lemos (2012), who averaged the governing equations (4)-(6) over volume and time. By contrast, only volume averaging was used in our derivation. By averaging Eqs. (4)-(6) in each REV and accounting for the zero mass flux at solid surfaces, we obtained the following macroscopic equations:

$$\frac{\partial \hat{u}_i}{\partial \hat{x}_i} = 0, \quad (8)$$

$$\frac{\partial \hat{u}_i}{\partial \hat{t}} + \frac{\partial (\hat{u}_i \hat{u}_j / \phi)}{\partial \hat{x}_j} + \frac{\partial (\phi \langle {}^i\tilde{u}_i {}^i\tilde{u}_j \rangle^i)}{\partial \hat{x}_j} = -\frac{\partial (\phi \langle \tilde{p} \rangle^i)}{\partial \hat{x}_i} + \frac{\text{Sc}}{\gamma_m \text{Ra}} \frac{\partial^2 \hat{u}_i}{\partial \hat{x}_j^2} - \frac{\phi \text{Sc}}{\gamma_m \text{RaDa}} z_i \hat{c} - \phi \hat{R}_i, \quad (9)$$

$$\frac{\partial (\phi \hat{c})}{\partial \hat{t}} + \frac{\partial (\hat{u}_i \hat{c})}{\partial \hat{x}_i} + \frac{\partial (\phi \langle {}^i\tilde{u}_i {}^i\tilde{c} \rangle^i)}{\partial \hat{x}_j} = \frac{1}{\text{Ra}} \frac{\partial^2 \hat{c}}{\partial \hat{x}_j^2}, \quad (10)$$

where ˆ denotes a volume averaged dimensionless quantity, $\phi$ is the porosity, and $\hat{u}_i = \phi \langle \tilde{u}_i \rangle^i$ and $\hat{c} = \frac{\langle c \rangle^i - c_0}{c_1 - c_0}$ are the volume averaged dimensionless velocity and species concentration, respectively. Here, the Rayleigh number for convection in porous media is defined as:

$$\text{Ra} \equiv \frac{\text{Ra}_f \text{Da}}{\gamma_m} = \frac{H \beta \Delta c g K}{D_m \nu}, \quad (11)$$

where $\gamma_m = \frac{D_m}{D_f}$ is the ratio of the effective mass diffusivity, $D_m$, and the fluid mass diffusivity, $D_f$. The effective mass diffusivity, $D_m$, is a macroscopic parameter which describes the mass diffusion through the porous medium. Further, note that the parameter $D_m$ is different from $D_f$ due to the effect of the porous matrix. Quantities $\phi\langle {}^i\tilde{u}_i {}^i\tilde{u}_j\rangle^i$ and $\phi\langle {}^i\tilde{u}_i {}^i\tilde{c}\rangle^i$ represent the momentum dispersion and species concentration dispersion, respectively.

The dimensionless total drag, $\hat{R}_i$, is usually modelled by the sum of the Darcy term and the Forchheimer term, i.e.,

$$\hat{R}_i = \frac{\text{Sc}}{\gamma_m \text{RaDa}} \hat{u}_i + \frac{C_F}{\text{Da}^{1/2}} |\hat{\mathbf{u}}| \hat{u}_i, \tag{12}$$

where $C_F$ is the (empirical) Forchheimer coefficient. Neglecting the higher order terms with respect to $Da = \frac{K}{H^2}$ and the dispersion terms $\frac{\partial(\phi\langle {}^i\tilde{u}_i {}^i\tilde{u}_j\rangle^i)}{\partial \hat{x}_j}$ and $\frac{\partial(\phi\langle {}^i\tilde{u}_i {}^i\tilde{c}\rangle^i)}{\partial \hat{x}_j}$, one obtains the traditional DOB equations:

$$\frac{\partial \hat{u}_i}{\partial \hat{x}_i} = 0, \tag{13}$$

$$\frac{\partial \hat{p}}{\partial \hat{x}_i} + \hat{c} z_i + \hat{u}_i = 0, \tag{14}$$

$$\frac{\partial \hat{c}}{\partial \hat{t}^*} + \frac{\partial (\hat{u}_i \hat{c})}{\partial \hat{x}_i} = \frac{1}{\text{Ra}} \frac{\partial^2 \hat{c}}{\partial \hat{x}_j^2}, \tag{15}$$

where $\hat{p} = \frac{\text{RaDa}\langle \tilde{p}\rangle^i}{\gamma_m \text{Sc}}$ is the normalized pressure. The dimensionless time is modified to be $\hat{t}^* = \hat{t}/\phi$. The Sherwood number for DOB studies is calculated as:

$$\text{Sh} = \frac{\int_w \overline{\frac{\partial \hat{c}}{\partial \hat{x}_2}} dA}{A_w}, \tag{16}$$

where $A_w$ is the surface area of the wall. Note that the definitions given by Eqs. (7) and (16) are equivalent and are in accordance with the definitions in previous DOB studies (Hewitt *et al.* 2012, 2013, 2014).

In the framework of the DOB equations (1)-(3), convection in porous media is uniquely determined by Ra, as defined in Eq. (11). Obviously, a prerequisite for the validity of the DOB equations is that the Darcy number is small in order for the terms of $O(1/Da)$ to dominate in Eq. (9). However, the Darcy number may affect the governing equations if Da is not small enough. In addition, convection in porous media may also be affected by the Schmidt number Sc, the porosity $\phi$, or pore-scale factors such as the pore scale $s$. The momentum and species dispersion terms may also influence the convection in porous media. Understanding whether these factors should be accounted for in the macroscopic equations requires further analysis.

2.3 Numerical method

A finite-volume method (FVM) was employed in the DNS. The solver was developed based on the open source code package OpenFoam 2.4. The solutions were advanced in time with the second-order implicit backward method. A second-order central-difference scheme was used for the spatial discretization. The pressure and velocity fields were corrected by the Pressure-Implicit scheme with Splitting of Operators (PISO) pressure-velocity coupling (Versteeg & Malalasekera 2007). A stabilized preconditioned (bi-)conjugate gradient solver was utilized to solve the pressure field and the momentum and species concentration equations.

A stream function method was used to solve the DOB equations (13)-(15). A stream function, $\psi$, was introduced for the velocity field, i.e.,

$$(\hat{u}_1, \hat{u}_2) = \left(\frac{\partial \psi}{\partial x_2}, -\frac{\partial \psi}{\partial x_1}\right). \tag{17}$$

The curl of the momentum equation, Eq. (14), gives:

$$\frac{\partial^2 \psi}{\partial x_i^2} = -\frac{\partial \hat{c}}{\partial x_1}. \tag{18}$$

In the stream function method, Eq. (15) is advanced in time to update the mass concentration field. Then, the Poisson equation, Eq. (18), is solved to calculate the stream function, $\psi$. Finally, the velocity is updated with Eq. (17). The second order implicit backward method was used for the time discretization. For the convection terms, we used linear interpolation to determine the variables at the cell interfaces. A second order central difference scheme was used for spatial discretization. A stabilized preconditioned (bi-)conjugate gradient solver was used to solve the Poisson equation (18). A preconditioned (bi-)conjugate gradient solver was used to solve the species concentration equations.

The code validation for our DNS solver was performed in our previous studies (Jin *et al.* 2015, 2017; Uth, *et al.* 2016). In these studies, as well as the present study, the finite volume method (FVM) was employed to simulate turbulent flows in channels with smooth and rough walls, and in porous media. The DNS results were compared with the DNS and experimental results reported in other references, as well as our DNS results obtained with the Lattice Boltzmann Method (LBM). The code validation for our DOB solver was performed in Kränzien & Jin (2019), where the obtained results were compared with Hewitt *et al.* (2012).

## 3. Studied test cases

The chosen two-dimensional porous matrix was composed of periodically arranged square obstacles of size *d*, which are a distance *s* apart in the lateral and vertical directions. The solid matrix was assumed to be adiabatic, i.e. the flux of species through the obstacles is zero. The porous matrix and the representative elementary volume (REV) are shown in figure 1(a). The porosity for the current porous matrix can be directly calculated from figure 1(a) as

$$\phi = 1 - \frac{d^2}{s^2}. \tag{19}$$

The porous matrix was composed of 800-20,000 REVs, and each REV was resolved by 1,600 to 6,400 mesh cells, yielding up to 72 million cells in the simulations. The same initial fields ($u_i = 0$ and $c = \frac{c_0+c_1}{2}$) were used in both DOB and DNS solutions. The largest DNS case was calculated using 384 processors in parallel for 960 hours.

The temporal evolution of the instantaneous Sherwood number Sh for a typical DNS is shown in figure 2. Time averaging was performed after Sh reached a statistically steady state, as indicated by the vertical dashed line. The computational time needed for this process depended on the flow parameters, such as Ra, *H/s*, and Sc.

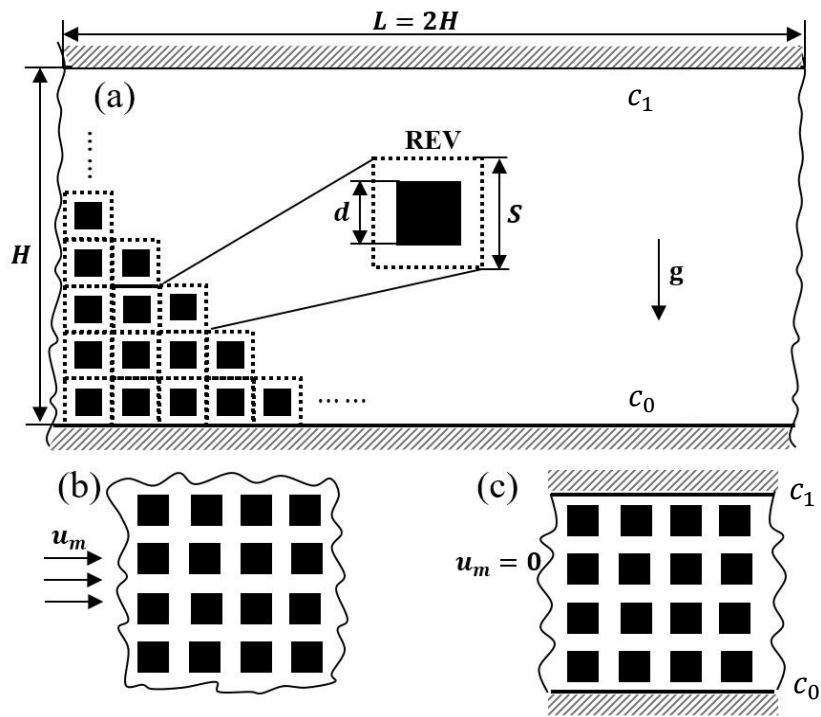

FIGURE 1. Porous matrices used in our DNS. In all cases, periodic boundary conditions are used in the horizontal direction. (a) Porous matrix for simulating convection in a porous medium. (b) Porous matrix for calculating the permeability $K$. (c) Porous matrix for calculating the effective mass diffusivity $D_m$. In (c), the mean velocity of the fluid is zero, thus mass transfer is via diffusion only.

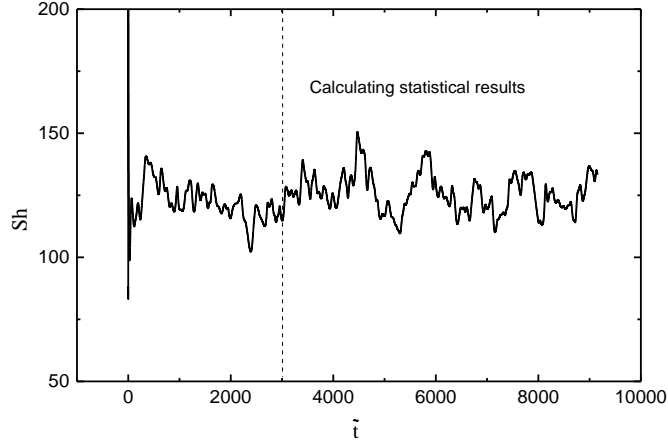

FIGURE 2. Sherwood number (Sh) versus dimensionless time $\tilde{t}$. Ra = 20,000, Sc = 250, and $H/s = 100$. The initial fields have $u_i = 0$ and $c = \frac{c_0+c_1}{2}$.

To compare the DNS results with the DOB results, the permeability, $K$, was estimated by simulating forced convection in the porous matrix shown in figure 1(b). Using this method, $K$ was calculated as the ratio of the applied pressure gradient to the mean velocity $u_m$. To determine the effective mass diffusivity, $D_m$, due to tortuosity, we simulated mass transfer in the same porous matrix, but bounded between two impermeable walls with $u_m = 0$ (see figure 1(c)). $D_m$ was calculated as $\frac{HD_f}{\Delta c A_w} \int_W \overline{\frac{\partial c}{\partial x_2}} dA$, where $A_w$ is the area of the upper (or lower) wall. Therefore, the Rayleigh number in our pore-resolved DNS study is the same as the one for the macroscopic (continuum scale) simulation. The values of the parameters used in our simulations are given in table 1.

We investigated cases characterized by Ra $\leq$ 20,000, scale ratios $H/s$ between 20 and 100, Schmidt numbers Sc = 1 and Sc = 250, and Darcy numbers between $2.8 \times 10^{-7}$ and $8.7 \times 10^{-6}$. According to the Kozeny's equation (Nield and Bejan 2017), the permeability $K$ can be approximated as $K^*$:

$$K^* = \frac{d^2 \phi^3}{\beta(1-\phi)^2}. \tag{20}$$

The pore size can be approximated from the permeability $K^*$ and the porosity $\phi$ as:

$$s^* = \left[\frac{\beta(1-\phi)K^*}{\phi^3}\right]^{1/2}, \tag{21}$$

where the empirical model coefficient $\beta = 126$ was used in this study. This value was obtained by fitting the values of $s^*$ to the real pore size $s$. The maximum difference between $s^*$ and $s$ was 4.8%, which we deemed an acceptable approximation.

Table 1. Main parameters for DNS and DOB cases.

| REV ID | $\phi$ | $s/d$ | $s^*/d$ | $K/d^2$ | $\gamma_m$ | Sc |
|---|---|---|---|---|---|---|
| a | 0.56 | 1.5 | 1.5 | 0.0079 | 0.38 | 1, 250 |
| b | 0.49 | 1.4 | 1.43 | 0.0042 | 0.32 | 1 |
| c | 0.36 | 1.25 | 1.31 | 0.0011 | 0.22 | 1, 250 |

We studied the sensitivity of the numerical results to the mesh resolution and time step for a test case with $H/s = 20$, $\phi = 0.56$ ($s/d = 1.5$), and Ra = 20,000 (see table 2). The numerical results showed that the Sherwood number is over-predicted when the mesh resolution is insufficient (case "f"), while it is under-predicted when the time step, indicated by the maximum Courant number ($\text{Co}_{\max}$), is too large (case "b"). The Sherwood numbers calculated for cases "a" and "c" through "e" vary by a maximum of 2.5%. We adopted this range of mesh resolutions and $\text{Co}_{\max}$ for all other test cases, including the cases with small Ra numbers. Thus, we estimate numerical errors in our DNS studies to be below 2.5%.

Table 2. Effects of the mesh resolution and maximum Courant number on the Sherwood number. The test case has the parameters $H/s = 20$, $\phi = 0.56$ ($s/d = 1.5$), and Ra = 20,000. The cases that are shaded grey were considered as converged (mesh and time step independent). $N_x$ and $N_y$ are the REV numbers in horizontal and vertical directions, while $N_{\text{REV}}$ is the number of mesh cells in each REV.

| Case ID | Mesh resolution ($N_x \times N_y \times N_{\text{REV}}$) | $\text{Co}_{\max}$ | Sh |
|---|---|---|---|
| a | $40 \times 20 \times 3600$ | 0.9 | 97.3 |
| b | $40 \times 20 \times 6400$ | 4.1 | 85.6 |
| c | $40 \times 20 \times 6400$ | 0.9 | 94.9 |
| d | $40 \times 20 \times 6400$ | 2.2 | 95.5 |
| e | $40 \times 20 \times 6400$ | 1.5 | 97 |
| f | $40 \times 20 \times 1600$ | 0.9 | 101.6 |

**4. Results and discussion**

4.1 Mega- and proto-plumes

A local Reynolds number $\text{Re}_K$ may be defined based on the permeability $K$ and the local velocity magnitude $|\mathbf{u}|$, i.e.,

$$\text{Re}_K = \frac{|\mathbf{u}|K^{1/2}}{\nu}. \tag{22}$$

Nield & Bejan (2017) indicated that the Darcy's term dominates the drag when $\text{Re}_K \ll 1$, while the Forchheimer's term has a greater effect on the flow for $\text{Re}_K > 1$. It should be noted that $|\mathbf{u}|$ in Eq. (22) cannot be approximated by the characteristic (macroscopic) velocity of the flow $u_m$, which is much larger than $|\mathbf{u}|$.

Figure 3 shows typical instantaneous fields of $\text{Re}_K$ for $\text{Ra} = 20{,}000$. The $\text{Re}_K$ values of all the cases for $\text{Sc} = 250$ are much smaller than unity and thus the flow is in the Darcy's regime. According to our DNS, the flow in the pores is laminar, despite the large Rayleigh number, because of the very small local Reynolds numbers ($\text{Re}_K$). The macroscopic flows at large Rayleigh numbers are nevertheless strongly nonlinear and chaotic, exhibiting "macroscopic turbulence". The convection for $\text{Sc} = 1$ is also generally in the Darcy's regime. The largest value of $\text{Re}_K$ for $\text{Sc} = 1$ is about 2.8, which occurs for the case of $\text{Ra} = 20{,}000$, $H/s = 20$, and $\phi = 0.56$. These results suggest that the Forchheimer's term may have more effect in heat transfer problems than in mass transfer problems because a Prandtl number of one is common in heat transfer, whereas typical Schmidt number values are larger than unity.

The instantaneous species concentration fields obtained from our DNS and DOB simulations at $\text{Ra} = 20{,}000$ are shown in figure 4. The macroscopic species concentration field $\langle c \rangle$ was either obtained from the DNS results by volume averaging over each REV (figure 4(a), (b)), or calculated directly in DOB simulations (figure 4(c)). Both DNS and DOB mass concentration fields exhibit two boundary layer regions and an interior region. The interior region is dominated by transient mega-plumes, whereas the boundary layers are filled with small proto-plumes. These small proto-plumes are products of the growing instabilities in the boundary layers that cause low concentration fluid to rise and high concentration fluid to sink.

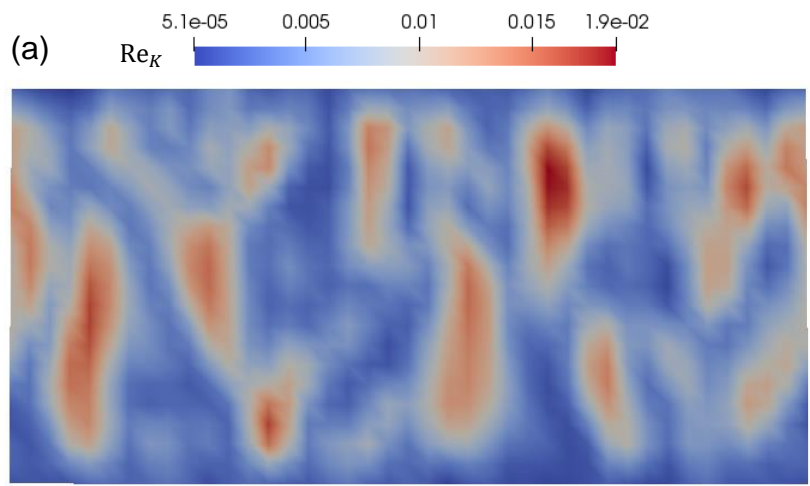

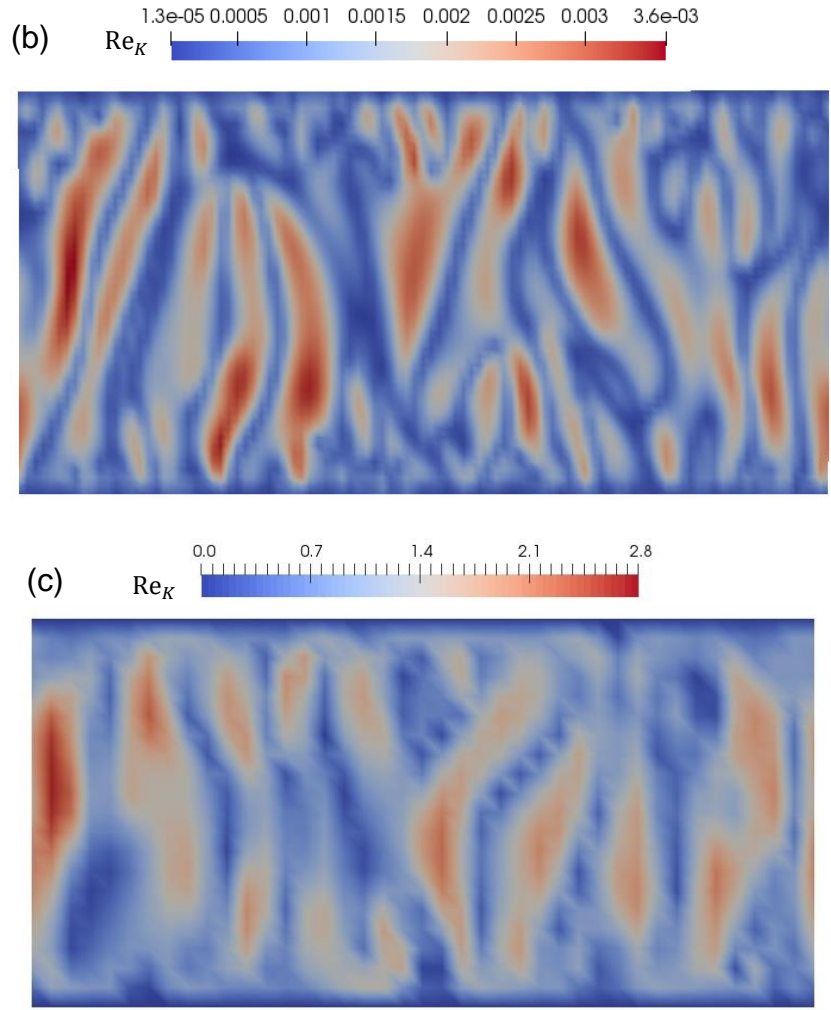

FIGURE 3. Snapshots of the instantaneous Reynolds number $Re_K$ based on the permeability at Ra = 20,000 from DNS results with $\phi$=0.56 ($s/d = 1.5$). (a) Sc = 250, $H/s = 20$; (b) Sc = 250, $H/s = 50$; (c) Sc = 1, $H/s = 20$.

Interestingly, the instantaneous species concentration fields obtained by DNS and displayed in figure 4(a) and 4(b) suggest that the size of mega- and proto-plumes increases with the pore size. This phenomenon is absent in classical Rayleigh-Bénard convection (without a porous medium) and cannot be captured by the DOB equations where pore effects enter the equations only via Ra.

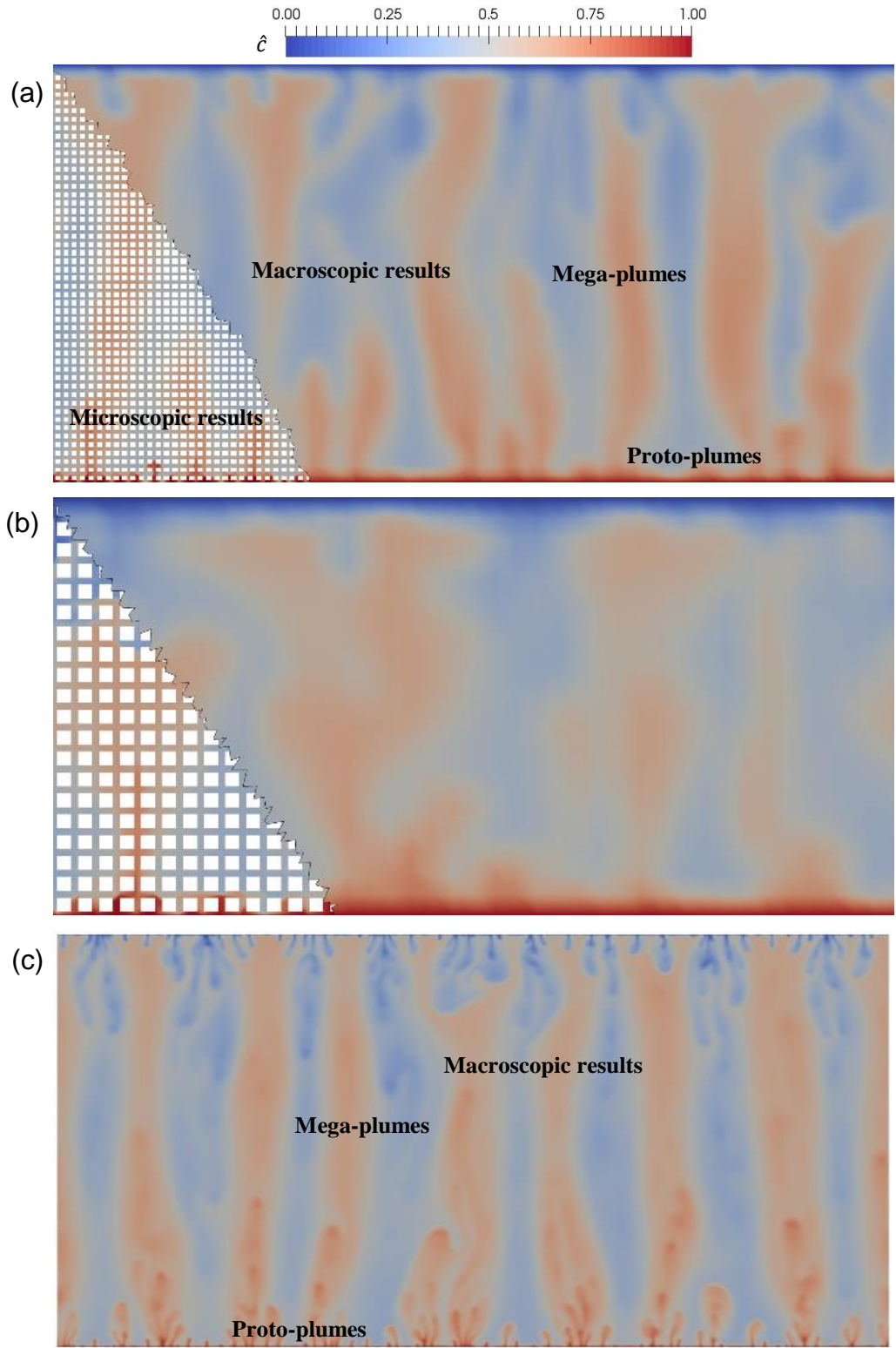

FIGURE 4. Snapshots of instantaneous mass concentration fields at Ra = 20,000 and Sc = 1. (a) DNS with $H/s$=50 and $\phi = 0.56$ ($s/d$=1.5). Here, the macroscopic mass concentration was obtained by volume averaging the microscopic mass concentration over each REV; (b) DNS with $H/s$=20 and $\phi = 0.56$ ($s/d$=1.5); (c) DOB simulation.

Figure 5 shows several instantaneous profiles of $\hat{c}$ along with their corresponding time-averaged discrete Fourier transforms at mid-height for Ra = 20,000 and Sc = 1. The mean concentration $\langle \hat{c} \rangle^{x1}$, where the operator $\langle \cdot \rangle^{x1}$ denotes horizontal averaging, was extracted from $\hat{c}$. The number of $\hat{c}$ maxima, which corresponds to the number of mega-plumes, decreases with increases in pore size $s$ (see figure 5(a)). In the DOB simulations, the peak amplitude occurs when the wavenumber $k$ equals $9\pi$. Wen *et al.* (2015) indicted that the wavenumber is not unique for Ra > 39,716. However, the wavenumber is well approximated by $k = 0.48\text{Ra}^{0.4}$ when Ra is smaller than this critical value. Figure 6 shows that our DOB results are close to this correlation, as well as to the DOB results of Hewitt *et al.* (2012) and Wen *et al.* (2015). A variability of the peak wavenumber due to long-time-scale fluctuations was observed in Hewitt *et al.* (2012) when different initial field or aspect ratio ($L/H$) were used, see hollow circles in figure 6. This variability was not found in our study since the same initial field and aspect ratio were used in both our DNS and DOB solutions.

In the DNS, the peaks are broader and the dominant wavenumber increases from $4\pi$ to $7\pi$ as $H/s$ increases from 20 to 50. This emphasizes the importance of the pore scale $s$ in shaping the mega-plumes. Motions at even larger length scales ($k \approx \pi$), which corresponds to the length of the box ($2H$), were observed in the DNS.

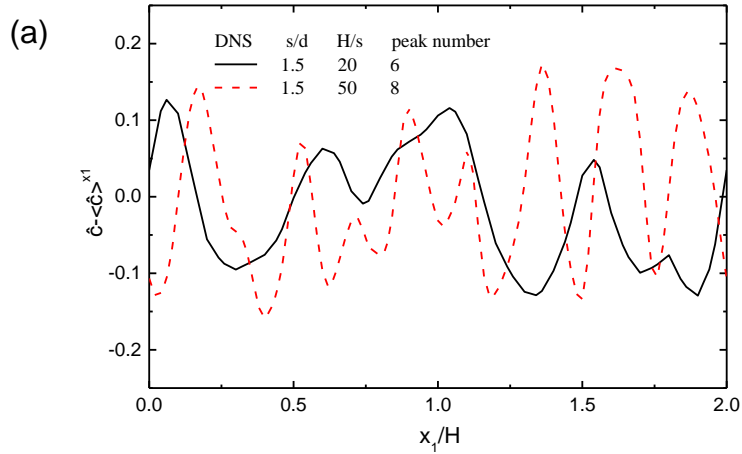

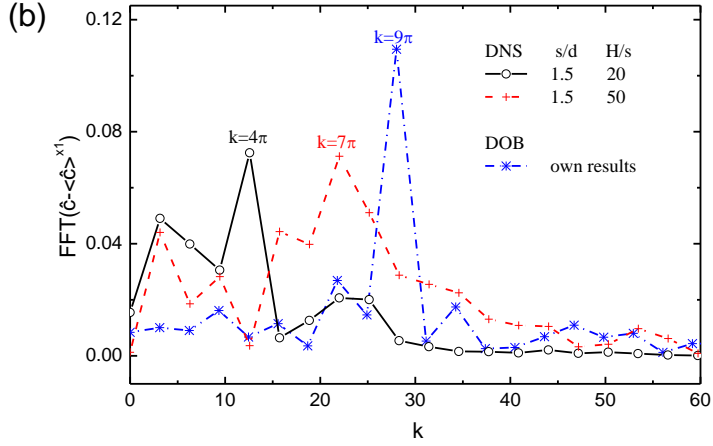

FIGURE 5. Instantaneous profiles (a) and average spectra (b) of the dimensionless mass concentration, $\hat{c}$, at mid-height $x_2/H = 0.5$. Ra = 20,000, $\phi = 0.56$ ($s/d$=1.5), and Sc = 1.

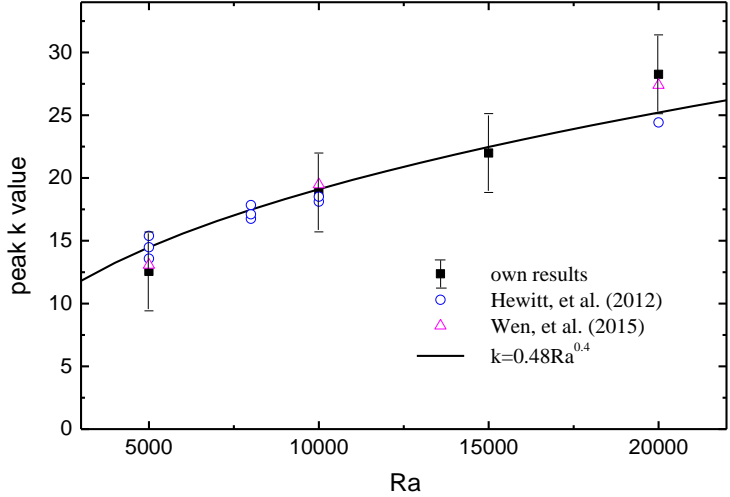

FIGURE 6. Peak wavenumber $k$ for the mega-plumes of the DOB results for different Rayleigh numbers. The maximum error bar for the peak wavenumber is estimated according to the wave number step size $\Delta k = \pi$ in our discrete Fast Fourier Transform (FFT).

Figure 7 shows several instantaneous profiles of $\hat{c}$ and their corresponding time-averaged Fourier transforms for Ra = 20,000 and Sc = 250. Similar to the results for Sc = 1, the dominant wavenumber increases from $4\pi$ to $7\pi$ as $H/s$ increases from 20 to 100. Here, the smallest Darcy number Da is $3.5 \times 10^{-7}$, which corresponds to $H/s = 100$. Large length scale ($k = \pi$) motions were also found in this test case, which makes the DNS results distinctly different from the DOB results. However, more detailed investigations are required to clarify the origin of these motions. We expect that the plumes will keep getting smaller as $H/s \rightarrow \infty$ and may even eventually converge toward the DOB results, see figure 8. However, although $Re_k$ for the DNS case is much smaller than 1 (see figure 3b for $H/s = 50$) and the pore size $s$ (0.01$H$ for $H/s = 100$) is much smaller than the plume size (about 0.23$H$ for Ra = 20,000, the DOB results), the peak wavenumbers from the DNS results are still clearly different from the DOB results. A possible reason is that no matter how small the pore size, the pore scale structure may always affect part of the boundary layer in the vicinity of the wall.

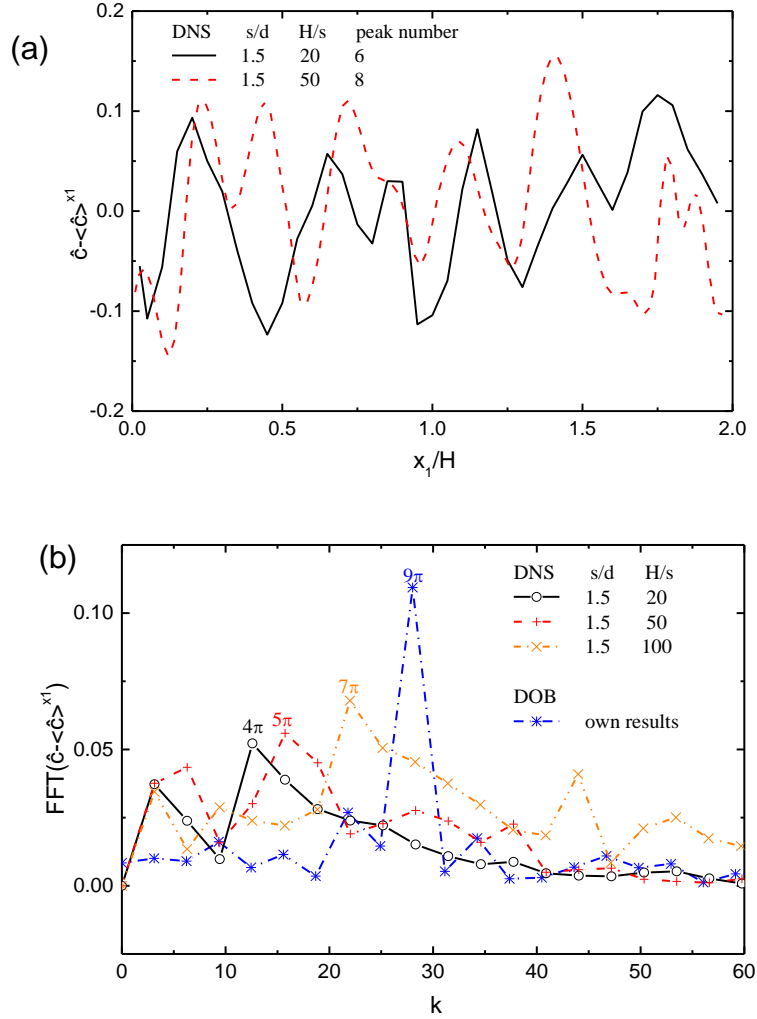

FIGURE 7. Instantaneous profiles (a) and average spectra (b) of the dimensionless mass concentration, $\hat{c}$, at mid-height $x_2/H = 0.5$, Ra = 20,000, $\phi = 0.56$ ($s/d$=1.5), and Sc = 250.

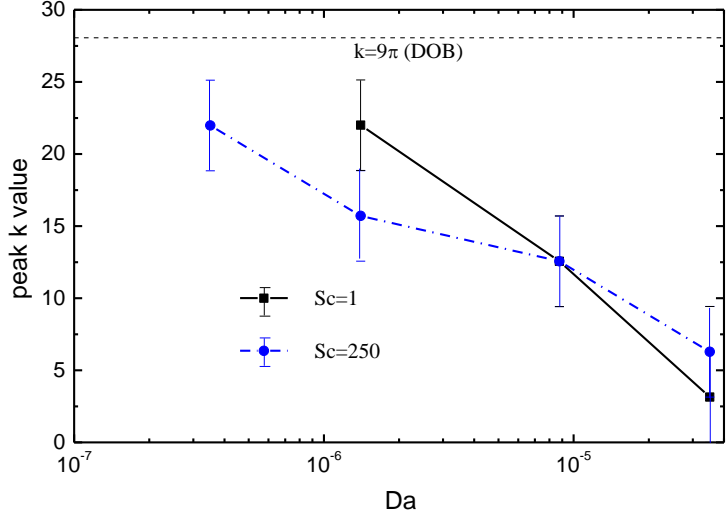

FIGURE 8. Peak wavenumber $k$ for the mega-plumes of the DNS results for different Darcy numbers, Ra = 20,000. The maximum error bar for the peak wavenumber is estimated according to the wave number step size $\Delta k = \pi$ in our discrete FFT.

4.2 Sherwood number

The scaling of the Sherwood number for $Sc = 1$ obtained from our DNS results, with various $H/s$ and porosity values, is shown in figure 9. The DOB simulations of Hewitt et al. (2012) and our own DOB results are compared with our DNS results (shown by solid and dashed lines in figure 8(a)). It appears that the DOB simulations overestimate the mass transfer rate for Ra in the range between 600 and 4,000, whereas at large Ra, they fall amidst the values obtained by DNS for the porosities studied here. Although the scale ratio $H/s$ appears to determine the scale of mega-plumes, it only mildly influences Sh and without a clear trend.

The DNS results show that the porosity has a significant effect on Sh in the high Rayleigh number regime (Ra > 10,000). For example, at Ra = 20,000, Sc = 1 and $H/s = 20$, Sh decreases from 158 to 96 while the porosity increases from $\phi = 0.36$ ($s/d = 1.25$) to 0.56 ($s/d = 1.5$). At large Rayleigh numbers (Ra > 5,000), the relationship between Sh and Ra can be well approximated by Sh = 8 + 0.0076Ra for $\phi$ =0.36. This linear relationship is in accordance with our own DOB results (Kränzien & Jin 2019), as well as the DOB results by Hewitt et al. (2012). However, for $\phi = 0.56$, the correlation Sh = $0.033\text{Ra}^{0.8}$ better fits the data.

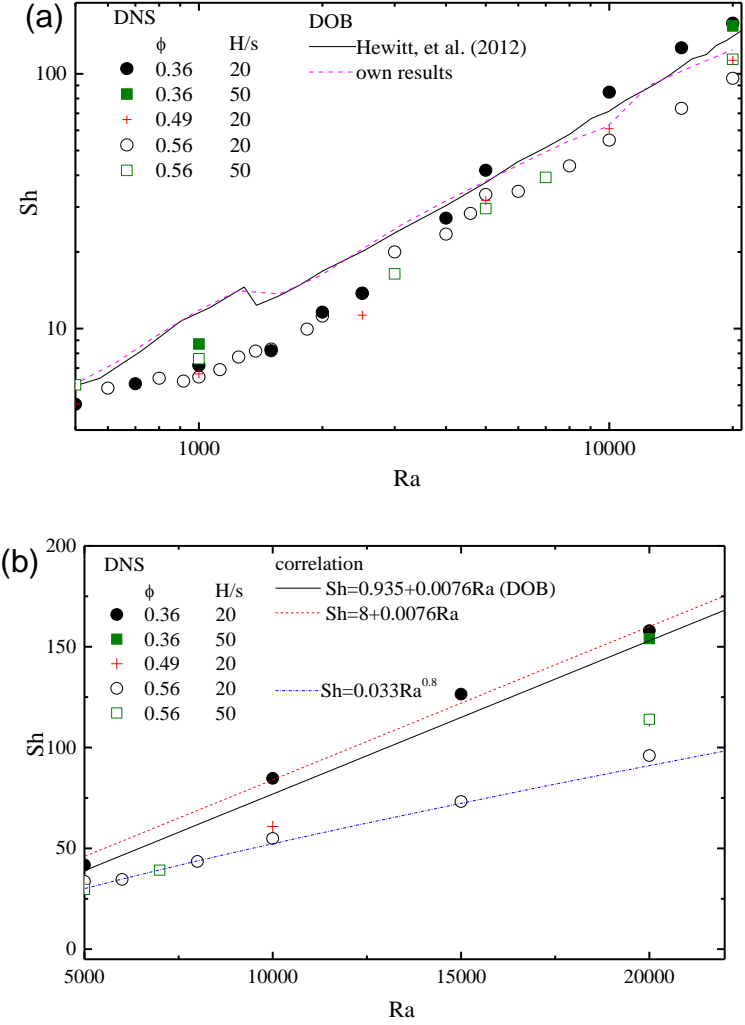

FIGURE 9. Time and surface averaged Sherwood number. Lines: DOB results; Symbols: DNS results. The porosity $\phi$ and scale ratio $H/s$ are varied. The results for $Sc = 1$ are shown in a loglog-scale (a), and a linear-scale for $Ra > 5,000$ (b).

The Sherwood number slightly increases as the Schmidt number is increased from 1 to 250. However, the relationship between Sh and Ra does not change qualitatively (see Fig .10). At large Rayleigh numbers ($Ra > 5,000$), the Sherwood number is still higher for a lower porosity. The $Sh = f(Ra)$ scaling changes from a linear scaling ($Sh = 16 + 0.0076Ra$) for $\phi = 0.36$ to a nonlinear one ($Sh = 0.045Ra^{0.8}$) for $\phi = 0.56$. The exponential coefficients for $Sc = 250$ are the same as those for $Sc = 1$ ($\alpha = 1$ for $\phi = 0.36$ and $\alpha = 0.8$ for $\phi = 0.56$), which indicates that the relationship between Sh and Ra does not change qualitatively if Sc is varied. Our DNS results suggest that the non-linear scaling laws found in the experiments by Backhaus *et al.* (2011) and Neufeld *et al.* (2010) may be related to the porosity or pore-scale factors, such as the pore size $s$.

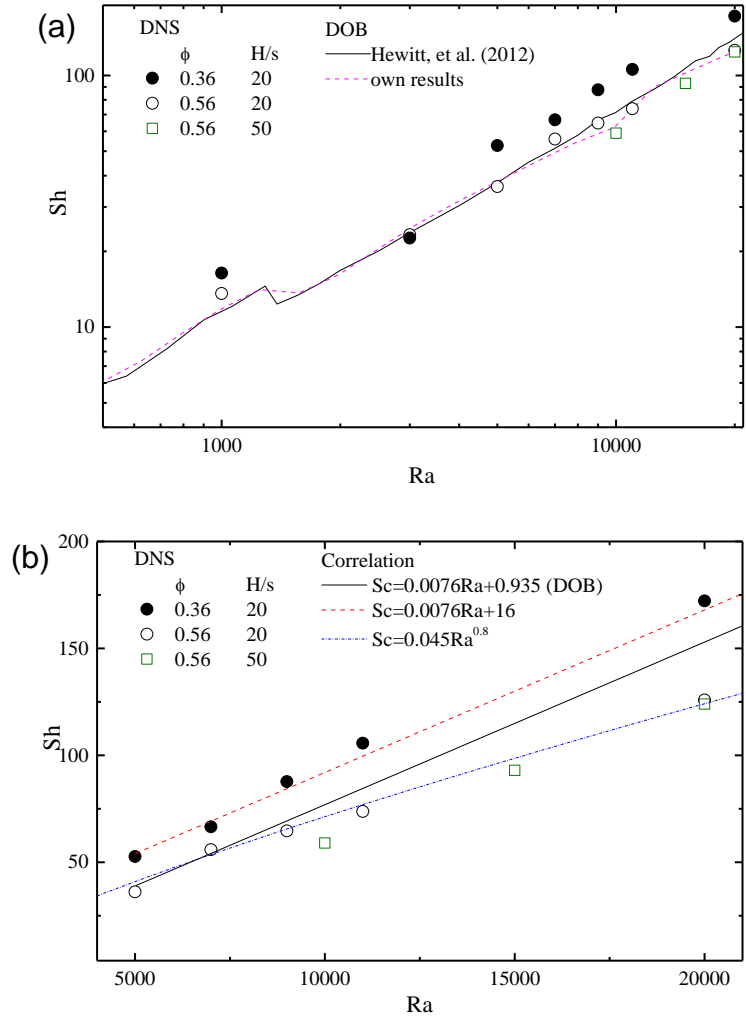

FIGURE 10. The time and surface (over the wall surface) averaged Sherwood number, $Sc = 250$. Lines: DOB results; Symbols: DNS results. The porosity $\phi$ and scale ratio $H/s$ are varied. The results are shown in the log-scale (a) and linear-scale for $Ra > 5,000$ (b).

It is worth noting that the small Schmidt number ($Sc = 1$) cases could also be representative of convective heat transfer in a porous medium with $Pr = 1$, which is common. However, in an experiment for heat transfer, Keene & Goldstein (2015) found a significantly smaller scaling exponent for the Nusselt number, $Nu \sim Ra^{0.319}$. The possible cause of this discrepancy is that conjugate heat transfer may play an important role, as we here assumed that the wall surfaces of the porous elements are adiabatic because we considered mass transfer. Clearly, the effect of conjugate heat transfer on convection in porous media deserves further investigation.

4.3 Mass concentration and velocity statistics

Figure 11 shows the vertical profiles of the temporally and horizontally averaged macroscopic mass concentration $\langle \bar{c} \rangle^{x1}$ obtained from DOB simulations. Here, the species concentration profiles at different Ra numbers become almost identical when the distance from the lower wall $\hat{x}_2$ is normalized by 1/Ra. This is in accordance with the statement by Huppert & Neufeld (2014) that the boundary layer thickness is determined by 1/Ra. More details of our DOB results can be found in Kränzien & Jin (2019).

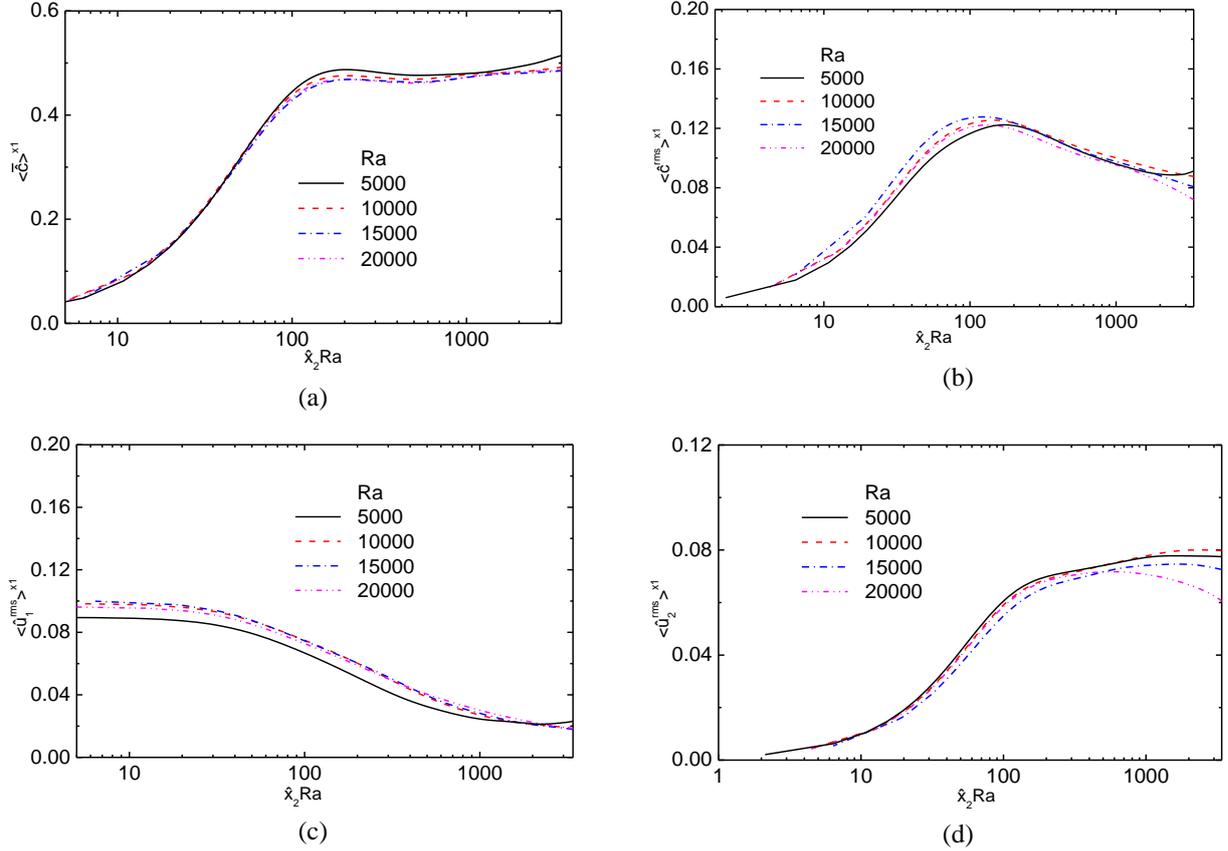

FIGURE 11. DOB results. (a) vertical profiles of time- and line-averaged species concentration; (b) r.m.s of species concentration fluctuation $\langle \hat{c}^{rms}\rangle^{x1}$; (c) $u_1$ fluctuation; (d) $u_2$ fluctuation.

By contrast, the profiles computed by resolving the flow at the pore scale with DNS exhibit a distinct length scale. Figure 12 shows strong deviations from the DNS results if $\hat{x}_2$ is scaled with 1/Ra. However, similar trends are observed if $\hat{x}_2$ is scaled with the pore scale $\hat{s} = s/H$ (figure 13). To explain this, one need to recall that the pore size $s$ can be approximated by using $s^*$, defined in Eq. (21) for a general two-dimensional porous matrix. The influence of the bounding walls is generally limited to within the first three REVs next to the walls, which leads to a steep gradient of $\langle \bar{c}\rangle^{x1}$ therein. When the distance from the lower wall is normalized by the pore size $s$, the $\langle \bar{c}\rangle^{x1}$ obtained at different Rayleigh numbers collapse together. Hence, the boundary layer thickness for convection in porous media is not determined by Ra, but by the pore size $s$.

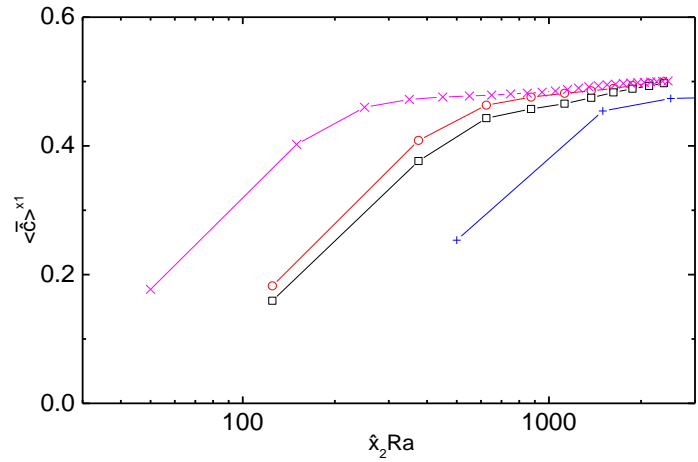

FIGURE 12 Vertical profiles of time- and line-averaged mass concentration for Sc = 1, DNS results. □: $H/s = 20$, $\phi = 0.36$, Ra = 5,000; ○: $H/s = 20$, $\phi = 0.56$, Ra = 5,000; +: $H/s = 20$, $\phi = 0.56$, Ra = 20,000; ×: $H/s = 50$, $\phi = 0.56$, Ra = 5,000.

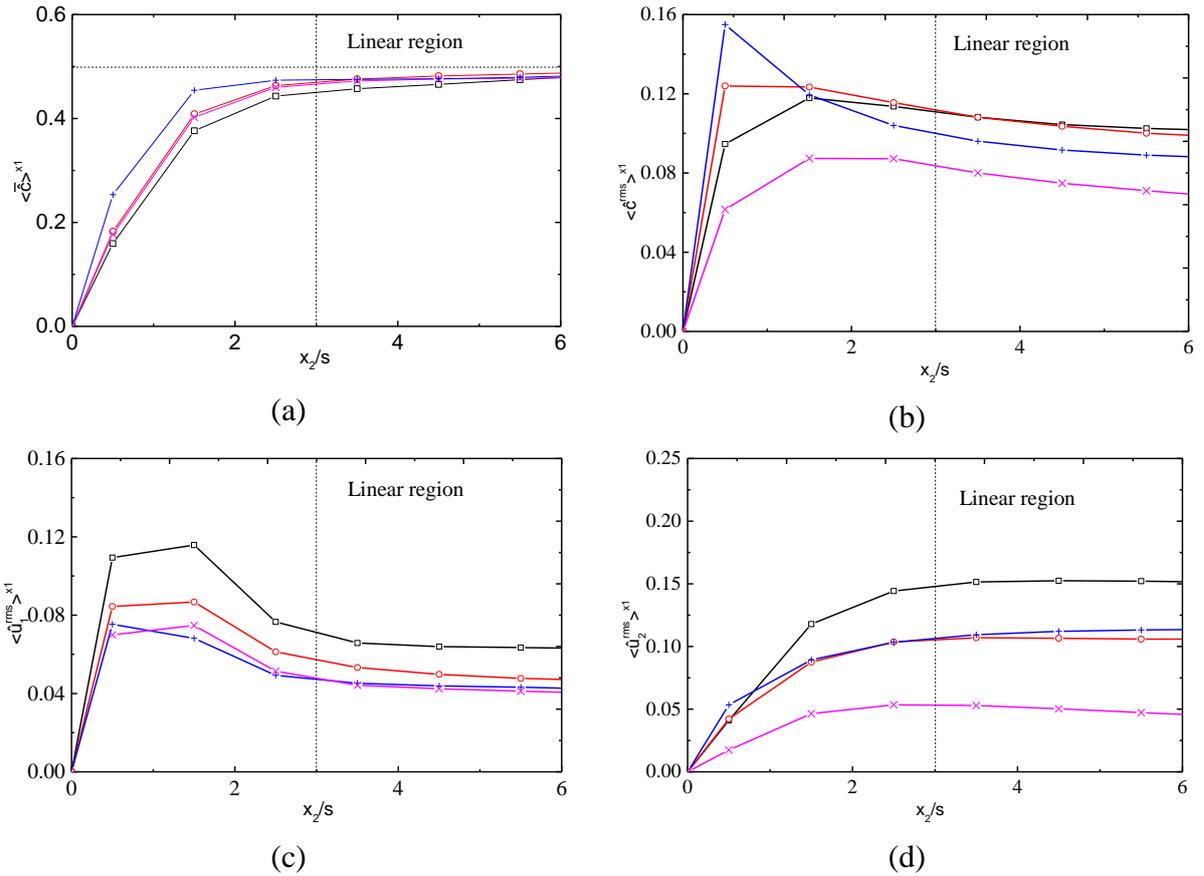

FIGURE 13. DNS results for Sc = 1. (a) Vertical profiles of time- and line-averaged mass concentration; (b) r.m.s of mass concentration fluctuation; (c) $u_1$ fluctuation; (d) $u_2$ fluctuation. □: $H/s = 20$, $\phi = 0.36$, Ra = 5,000; ○: $H/s = 20$, $\phi = 0.56$, Ra = 5,000; +: $H/s = 20$, $\phi = 0.56$, Ra = 20,000; ×: $H/s = 50$, $\phi = 0.56$, Ra = 5,000.

Figure 13 also shows the root-mean-square (r.m.s.) macroscopic mass concentration and velocity fluctuations, $\langle \hat{c}^{rms} \rangle^{x1}$, $\langle \hat{u}_1^{rms} \rangle^{x1}$, and $\langle \hat{u}_2^{rms} \rangle^{x1}$, respectively, for Sc = 1. The velocity fluctuations were normalized with the characteristic velocity, $u_m = \beta \Delta c g K / \nu$. The DNS results show that, besides Ra, the scale ratios $H/s$ and porosity $\phi$ also have significant effects on r.m.s. quantities. These effects are absent in DOB simulations. In addition, $\langle \hat{u}_1^{rms} \rangle^{x1}$ in figure 13(b) is distinctly different from the DOB results, which have unphysical (non-zero) velocity fluctuations at the wall surface (Hewitt, et al. 2012; Kränzien & Jin 2019). A possible reason for these discrepancies is that momentum dispersion is not accounted for in the DOB equations.

Figure 14 shows the temporally and horizontally averaged macroscopic results for Sc = 250. Similar to the results for Sc = 1 (see Fig. 13), the influence of the bounding walls is still limited to within the first three REVs next to the walls. Again, our DNS results confirm that the boundary layer thickness for convection in porous media is determined by the pore size $s$ (or $s^*$ for a porous matrix with permeability $K$ and porosity $\phi$).

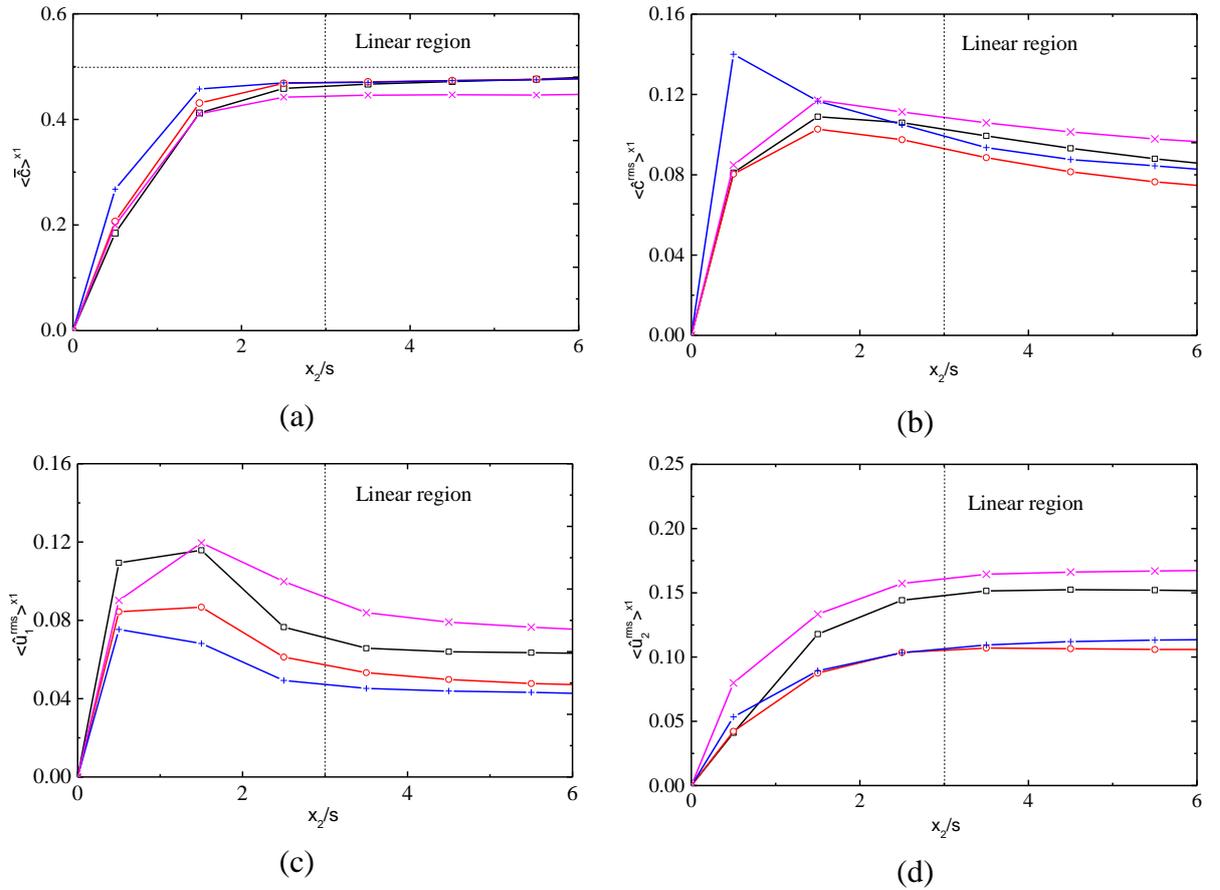

FIGURE 14. DNS results for Sc = 250. (a) Vertical profiles of time- and line-averaged mass concentration; (b) r.m.s of mass concentration fluctuation; (c) $u_1$ fluctuation; (d) $u_2$ fluctuation. □: $H/s = 20$, $\phi = 0.36$, Ra = 5,000; ○: $H/s = 20$, $\phi = 0.56$, Ra = 5,000; +:

$H/s = 20$, $\phi = 0.56$, Ra = 20,000; ×: $H/s$=50, $\phi = 0.56$, Ra = 20,000.

As the boundary layer thickness is determined by the pore size (which can be characterized by the square root of the permeability, $\sqrt{K}$), the momentum dispersion term $\frac{\partial(\phi\langle^i\tilde{u}_i{}^i\tilde{u}_j\rangle^i)}{\partial\hat{x}_j}$ and viscous diffusion term $\frac{\text{Sc}}{\gamma_m \text{Ra}}\frac{\partial^2 \hat{u}_i}{\partial \hat{x}_j^2}$ in the macroscopic equation (Eq. 9) are expected to scale as $1/K$. Thus, the momentum dispersion and viscous diffusion terms should be of order $\frac{1}{Da} = \frac{H^2}{K}$, exactly as the Darcy term itself. Thus, our DNS results suggest that the pore scale significantly influences convection in porous media through the momentum dispersion and viscous diffusion terms, which are neglected in the DOB equations. We conclude that these terms cannot be neglected in the macroscopic equations even if the pore size is small. In addition, the pore size should be used as the characteristic length when the dispersion term is modeled.

## 5. Conclusions

Natural convection in a porous medium, made of two-dimensional square obstacles, was studied with DNS by fully resolving the flow field within the pores. Upon comparing our DNS and DOB results, we found significant effects of the pore-scale on the macroscopic flow and mass concentration fields. These effects are summarized in what follows.

The boundary layer thickness is not determined by 1/Ra, as DOB simulations suggest (Huppert & Neufeld 2014), but by the pore size. In addition, the sizes of the mega-plumes in the interior region increase with increasing pore size. The DNS exhibit motions with even larger length scales, reaching up to the domain size. This is different from the DOB simulations, where the sizes of mega-plumes in the interior region depend only on the Rayleigh number. Furthermore, note that the spectra of the DNS exhibit much broader peaks and many secondary peaks at larger and smaller wavenumbers, which are entirely missing in the DOB case. Hence, even if the dominant wavenumber appears to converge to the DOB case, it is unclear whether the entire spectrum will converge to it. Overall, pore-scale effects at a low Schmidt number (Sc=1) are qualitatively similar to those at a high Schmidt number (Sc=250).

The porosity was found to have a strong impact on the mass transfer. At high Rayleigh numbers, increasing the porosity resulted in a lower Sherwood number. More importantly, the Sh = $f$(Ra) relationship changed from a linear scaling law (Sh~Ra) for $\phi = 0.36$ to a nonlinear scaling law (Sh~Ra$^{0.8}$) for $\phi = 0.56$. This is in accordance with the study of Bernard-Rayleigh flows without a porous medium. For example, Shishkina & Sebastian (2016) suggested the scaling Nu $\sim$ Ra$^{1/4}$ for large Pr and Ra numbers. This is also observed in the experiments of Backhaus et al. (2011) and Neufeld et al. (2010). However, this comparison must be taken with caution because the geometrical details of porous matrices used in the experiments are not specified in their studies.

One limitation of our results is the relatively small $H/s \leq 100$ values used in the DNS (to keep computational costs manageable), whereas problems such as $CO_2$ sequestration are characterized by very large $H/s$ ratios. However, the $H/s$ values in our study ensured that most cases (all cases for Sc=250) were within the Darcy's regime. Therefore, $H/s$ only has a mild influence on Sh without a clear trend within the range of our computational parameters. This suggests that DNS with reduced *H/s* values may be used for tackling practical problems such as $CO_2$ sequestration despite a much larger *H/s* ratio in that situation.

We stress that none of the effects revealed here can be captured by state-of-the-art DOB simulations, where all the microscopic details of the porous media are lumped into the Rayleigh number. Our simulations open avenues for the extension and parametrization of improved DOB equations including non-Darcy terms, accounting for the effect of porosity and the pore scale. More specifically, the momentum dispersion and viscous diffusion terms should be accounted for in the macroscopic equations, even when the pore size is much smaller than the plume size and $\text{Re}_k \ll 1$. The pore size should be used as the characteristic length when the dispersion term is modeled. To achieve this goal, more extensive numerical and experimental studies involving more realistic porous matrices to predict $CO_2$ sequestration are needed.

**Acknowledgments**

The authors gratefully acknowledge the support of this study by the DFG (Deutsche Forschungsgemeinschaft) and the HLRN (North-German Supercomputing Alliance). AVK acknowledges with gratitude the support of the National Science Foundation (award CBET-1642262) and the Alexander von Humboldt Foundation through the Humboldt Research Award.

Declaration of Interests. The authors report no conflict of interest.